\documentclass{webofc}
\usepackage[varg]{txfonts}   
\usepackage{color}
\newcommand{\citen}[1]{\citep*{#1}}
\newcommand{\taumodel}{$\tau$-model}
\newcommand{\adhoc}{\textit{ad hoc}}

\newcommand{\etal}{{\it et~al.}}%
\newcommand{\eg}{{\it e.g.}}%
\newcommand{\ie}{{\it i.e.}}%
\newcommand{\Eq}[1]{Eq.\,(\ref{#1})}%
\newcommand{\Eqs}[1]{Eqs.\,(\ref{#1})}%
\newcommand{\Fig}[1]{Fig.\,\ref{#1}}%
\newcommand{\Tab}[1]{Table~\ref{#1}}%
\newcommand{\Pep}{e$^+$}%
\newcommand{\Pem}{e$^-$}%
\newcommand{\Pee}{\Pep\Pem}%
\newcommand{\PZ}{\ensuremath{\mathrm{Z}}}
\newcommand{\Pq}{\ensuremath{\mathrm{q}}}
\newcommand{\Paq}{\ensuremath{\bar{\mathrm{q}}}}
\newcommand{\Pg}{\ensuremath{\mathrm{g}}}
\newcommand{\eV}{\hbox{\ensuremath{\mathrm{e\kern-0.1em V}}}}%
\newcommand{\GeV}{\hbox{\ensuremath{\mathrm{G}}\eV}}%

\newcommand{\abs}[1]{\left|#1\right|}                        
\newcommand{\chisq}{\ensuremath{\chi^2}}
\newcommand{\ycut}{\ensuremath{y_\mathrm{cut}}}
\newcommand{\mt}{\ensuremath{m_\mathrm{t}}}
 
\newcommand{\Ra}{\ensuremath{R_\mathrm{a}}}

\graphicspath{{./}}     

\begin{document}
\title{{Interpreting BEC in \Pee\ annihilation}
        \footnote{Talk given at XLVIII International Symposium on Multiparticle Dynamics, Singapore, 3--7 Sep. 2018}}
%
%
 
\author{\firstname{W.J.} \lastname{Metzger}\inst{1}\fnsep\thanks{\email{W.Metzger@science.ru.nl}}
        \and
        \firstname{T.} \lastname{Cs\"org\H{o}}\inst{2,3}
        \and
        \firstname{T.} \lastname{Nov\'ak}\inst{3}
        \and
        \firstname{S.} \lastname{L\"ok\"os}\inst{3,4}
}
 
\institute{IMAPP, Radboud University, NL-6525 AJ\ \ Nijmegen, The Netherlands
\and
          {Wigner RCP, Konkoly-Thege 29-33, H-1121 Budapest XII, Hungary}
\and
          {Eszterh{\'a}zy University, M{\'a}trai {\'u}t 36, H-3200 Gy{\"o}ngy{\"o}s, Hungary}
\and
          {E{\"o}tv{\"o}s University, P{\'a}zm{\'a}ny P{\'e}ter S{\'e}t{\'a}ny 1/A, H-1111 Budapest, Hungary}
          }
 
\abstract{%
  The usual interpretation of Bose-Einstein correlations (BEC) of identical boson pairs
relates the width of the peak in the correlation function at small relative four-momentum
to the spatial extent of the source of the bosons.  However, in the \taumodel, which
successfully describes BEC in hadronic \PZ\ decay, the width of the peak is related to
the temporal extent of boson emission.  Some new checks on the validity of both the \taumodel\
and the usual descriptions are presented.
}
\maketitle
\section{Introduction}\label{intro}
First a brief review of `classic' parametrizations of Bose-Einstein Correlations (BEC) is given and contrasted with the
parametrization of the \taumodel~\cite{Tamas;Zimanji:1990,ourTauModel}.
which has been found~\cite{L3_levy:2011} to describe well
Bose-Einstein correlations in hadronic \PZ\ decay.
 
The data used in this paper are taken from  Ref.~\citen{L3_levy:2011}. They comprise both two-jet and three-jet events,
as determined using the Durham jet algorithm~\cite{durham,durham2,durham3} with resolution parameter \ycut=0.006,
from \Pee\ annihilation at the \PZ-pole.

\subsection{`Classic' Parametrizations}    \label{classic}
The    Bose-Einstein correlation function, $R_2$,
is measured by $R_2(Q)=\rho(Q)/\rho_0(Q)$,
where $\rho(Q)$ is the density of identical boson pairs with invariant four-momentum difference
$Q=\sqrt{-(p_1-p_2)^2}$
and $\rho_0(Q)$ is the similar density in an artificially constructed reference sample,
which should differ from the data only in that it does not contain the effects of Bose symmetrization of identical bosons.
It is often parametrized as
\begin{equation} \label{eq:R2parametriz}
  R_2 =    \gamma \left[ 1+ \lambda G \right] (1+ \epsilon Q) \;,
\end{equation}
with
\begin{equation} \label{eq:gauss_param}
  G   =                             \exp \left(-\left(rQ\right)^{2} \right) \;.
\end{equation}
The corresponding distribution of boson emission points in space-time is a spherically symmetric Gaussian with
standard deviation $r$.
 
The factor $(1+\epsilon Q)$ is included to account for non-BEC which are not removed by $\rho_0$, \ie,
to make up for slight inadequacies in $\rho_0$, and $\gamma$ is a normalization parameter.
The parameter $\lambda$ is introduced to account for effects reducing the
amount of BEC, \eg, some of the bosons coming from resonance decays, or the identical bosons being partially coherent.
 
However, this description was found~\cite{L3_levy:2011} not to describe the \PZ-decay data, a better description being provided by
the Edgeworth expansion about the Gaussian~\cite{L3_3D:1999}:
\begin{equation} \label{eq:edgew_param}
  G   =                             \exp \left(-\left(rQ\right)^{2} \right)
           \cdot\left[1+\frac{\kappa}{3!} H_{3}(rQ)\right]  \;,
\end{equation}
where $H_3$ is the third-order Hermite polynomial.
 
A different way to depart from the Gaussian is the generalization to a symmetric L\'evy stable distribution. Then
\begin{equation} \label{eq:symlevy_param}
  G   =                             \exp \left(-\left(rQ\right)^{\alpha} \right) \:,
\end{equation}
where $0<\alpha\le2$ is the so-called index of stability, which was introduced to BEC in Ref.~\cite{Tamas:Levy2004}.

\begin{figure}[h]
  \mbox{\includegraphics[width=.45\textwidth,clip,viewport=56 87  518 680]{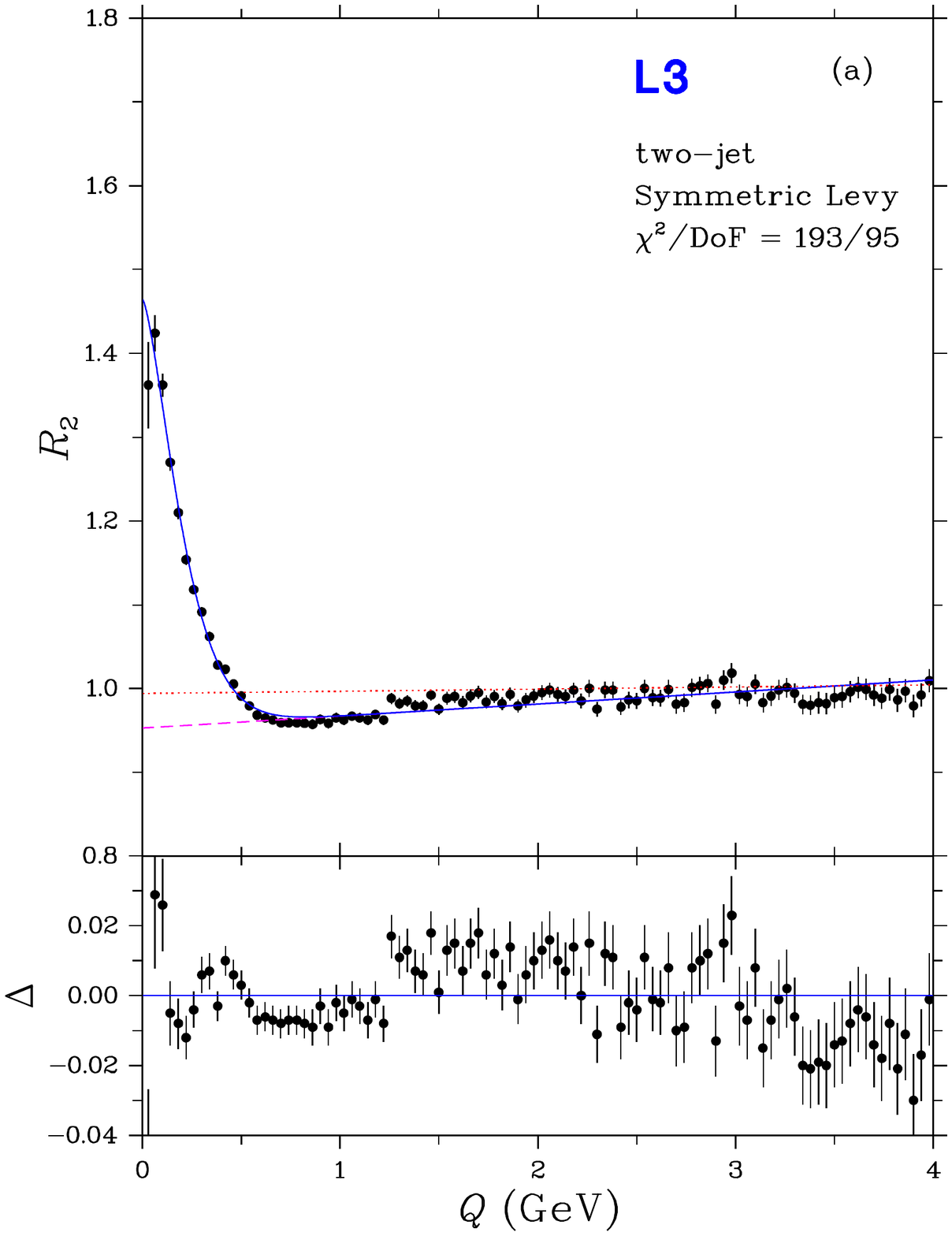}}
  \mbox{\includegraphics[width=.45\textwidth,clip,viewport=56 87  518 680]{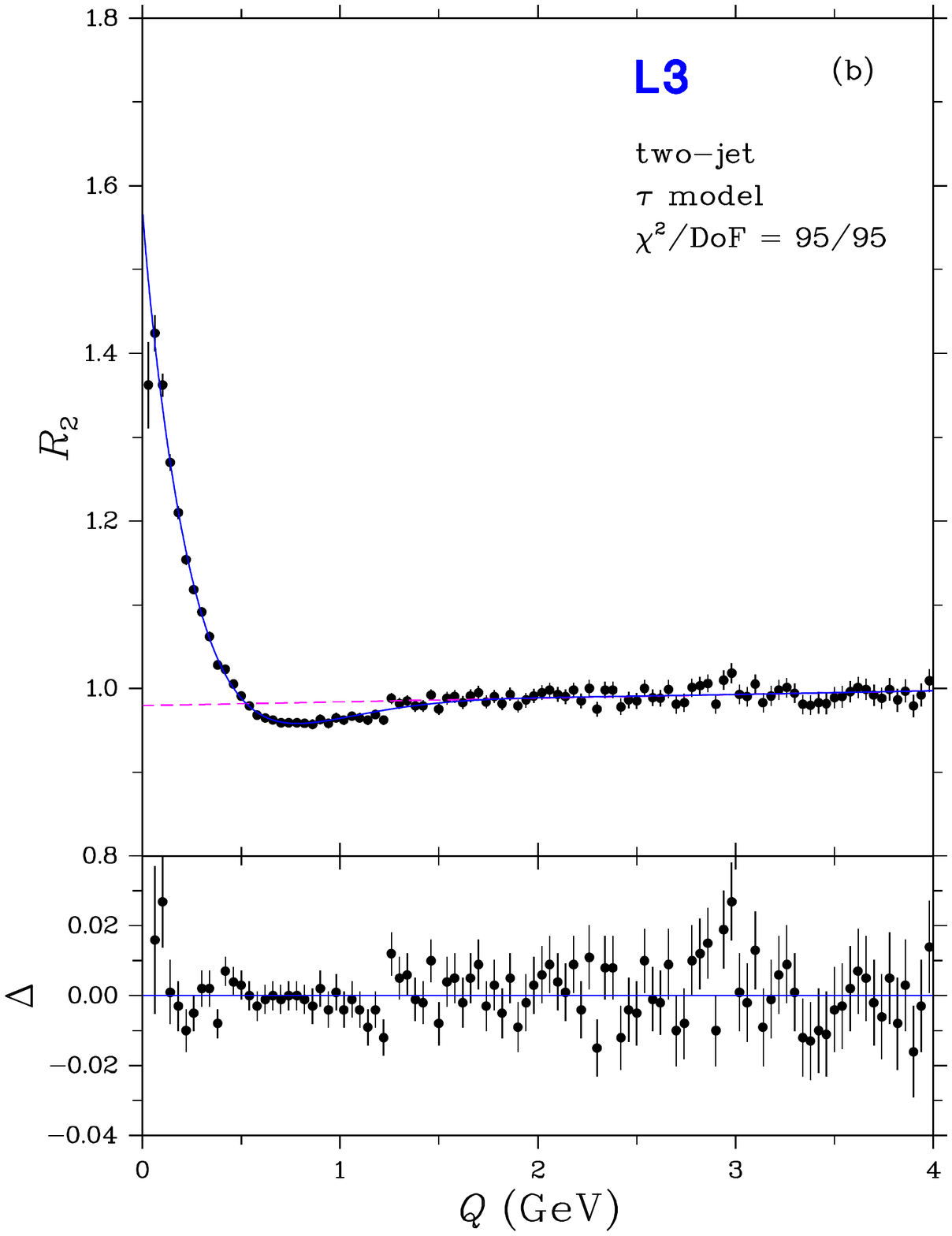}}
        \caption{(a) Symmetric L\'evy fit for two-jet events;
                 (b) Simplified \taumodel\ fit for two-jet events.
        }
        \label{fig:symlevy2jet}
\end{figure}

A fit of the Edgeworth parametrization to the two-jet data of Ref.~\citen{L3_levy:2011} finds
$\kappa=0.71\pm0.06$, while a fit of the symmetric L\'evy parametrization yields $\alpha=1.34\pm0.04$.
Both values are far from the corresponding Gaussian values of $\kappa=0$ and $\alpha=2$, respectively.
Although the \chisq\ of these fits are a great improvement over that of the Gaussian fit, they are still very large.
The corresponding confidence levels are $10^{-15}$ for the Gaussian and $10^{-5}$ and $10^{-8}$ for the Edgeworth and L\'evy
fits, respectively.   The symmetric L\'evy fit is shown in \Fig{fig:symlevy2jet}a.

The reason for the failure of the `classic' parametrizations is readily apparent from
\Fig{fig:symlevy2jet}a. There is a region of anti-correlation ($R_2<1$) extending from about $Q=0.5$ to $1.5\,\GeV$.
The `classic' parametrizations, which are of the form $1+P$, where $P$ is a positive-definite quantity, are unable
to accomadate the anti-correlation.
This was not realized for a long time because experiments only plotted the correlation function up to $Q=2\;\GeV$
or less.  In Ref.~\citen{L3_levy:2011} $Q$ was plotted to 4\,\GeV, and the
anti-correlation became apparent. This anti-correlation,
which one might term Bose-Einstein Anti-Correlations (BEAC),
as well as the Bose-Einstein correlations (BEC) are both well described by the \taumodel.

\subsection{The \taumodel}    \label{taumodel}
The \taumodel~\cite{Tamas;Zimanji:1990,ourTauModel}
rests on several assumptions:
\begin{itemize}
   \item The average production point is proportional to the momentum of the emitted boson.
     Dimensionally the momentum must be multiplied by a time divided by a mass to yield a spatial dimension.
     The description of a two-jet event is invariant to Lorentz boosts along the direction of the colour field.
     The relevant boost-invariant quantities are the ``longitudinal'' proper time,
     $\tau=\sqrt{\overline{t}^2-\overline{r}^2_z}$
     and the ``transverse'' mass,
     $\mt=\sqrt{E^2-p_z^2}$, resulting in
     \begin{equation} \label{eq:xpcorr}
        \overline{x}^\mu (p^\mu)  =  a \tau\ p^\mu \;, \qquad a=1/\mt \;.
     \end{equation}
   \item The spatial distribution of production points about their mean is very narrow,
     although the distribution of proper time may be broad.
   \item The distribution of $\tau$ is a one-sided L\'evy distribution,
     one-sided because no particles are emitted before the \Pee\ collision.
\end{itemize}
Then $R_2$ turns out to depend on three variables, $Q$ and the transverse mass of each of the particles making up the pair:
\begin{eqnarray}
   R_2 (Q, a_1, a_2) &=&
       \gamma
       \left\{ 1+\lambda
      \cos\left[\frac{\tau_0 Q^2 (a_1+a_2)}{2} +
           \tan\left(\frac{\alpha\pi}{2}\right)
           \left(\frac{\Delta\tau {Q^2}}{2}\right)^{\!\alpha}
           \frac{a_{1}^{\alpha}+a_{2}^\alpha}{2} \right]
  \right. \nonumber  \\
     & & \hfill
  \left.     \cdot
      \exp\left[-\left(\frac{\Delta\tau {Q^2}}{2}\right)^{\!\alpha}
          \frac{ a_{1}^{\alpha}+a_{2}^{\alpha}}{2} \right]
      \right\}
        \cdot\left(1+\epsilon Q\right) \;.
  \label{eq:R2taumodel}
\end{eqnarray}
 
Fits in three dimensions are problematic with the available statistics. Hence we simplify this expression by
introducing an effective radius, $R$, defined by
     \begin{equation} \label{eq:effR}
         R = \left(\frac{\Delta\tau}{2}\right)^{\!\alpha}
               \frac{a_{1}^{\alpha}+a_{2}^{\alpha}}{2}      \;.
     \end{equation}
Further, we assume that particle production begins immediately, \ie, $\tau_0=0$.
Then
%
\begin{subequations} \label{eq:asymlevR2}
\begin{align}
    R_2(Q) &= \gamma \left[ 1+ \lambda \cos \left(\left(\Ra Q\right)^{2\alpha} \right)
             \exp \left(-\left(RQ\right)^{2\alpha} \right) \right] (1+ \epsilon Q) \;, \label{eq:asymlevR2_}   \\
    R_\mathrm{a}^{2\alpha} &= \tan\left(\frac{\alpha\pi}{2}\right) R^{2\alpha} \;.     \label{eq:asymlevRaR}
\end{align}
\end{subequations}
The fit of the simplified \taumodel\ to the two-jet data is shown in \Fig{fig:symlevy2jet}b.
Unlike the fits of the classic parametrizations, the \chisq\ is acceptable, and the residuals lack structure.

Note that the difference between the parametrizations of \Eqs{eq:symlevy_param} and (\ref{eq:asymlevR2})
is the presence of the cosine term, which provides the description of the BEAC dip.
The parameter $R$ describes the BEC peak, and \Ra\  describes the anti-correlation region.
While one might have had the insight to add, \adhoc, a $\cos$ term to \Eq{eq:symlevy_param}, it is the \taumodel\
which provides a physical reason for it and
which predicts a relationship,     \Eq{eq:asymlevRaR}, between $R$ and \Ra, \ie, between
the correlation and the anti-correlation.

\section{Expansions} \label{sect:expansion}
Recall that the Edgeworth expansion of the Gaussian parametrization provided evidence (in addition to the poor \chisq)
that the Gaussian was inadequate.
In this section we look at expansions of the Symmetric L\'evy and the \taumodel\ parametrizations.
 
\subsection{Symmetric L\'evy}
\begin{figure}[h]
   \mbox{\includegraphics[width=.45\textwidth,clip,viewport=56 87  518 680]{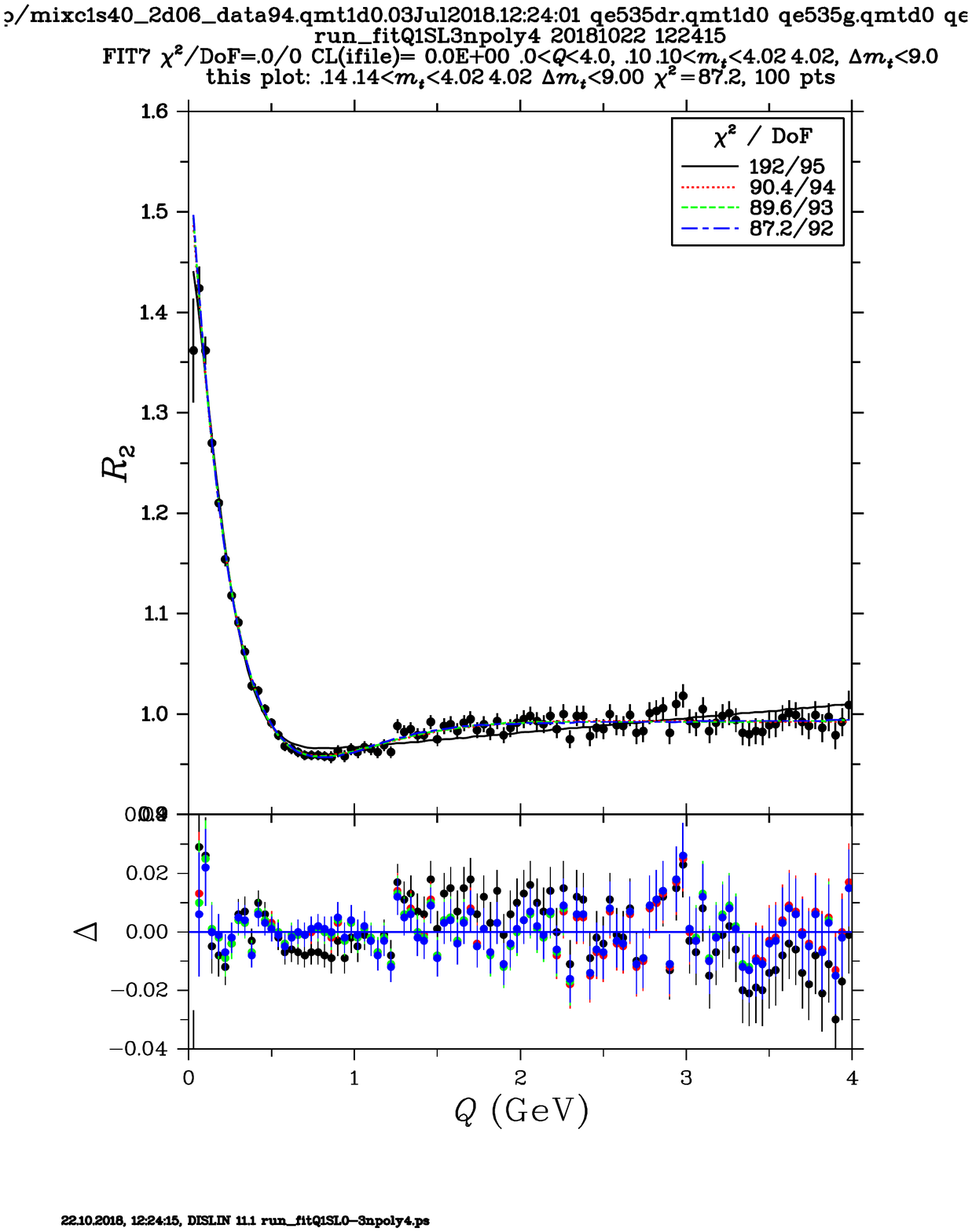}}
   \mbox{\includegraphics[width=.45\textwidth,clip,viewport=56 87  518 680]{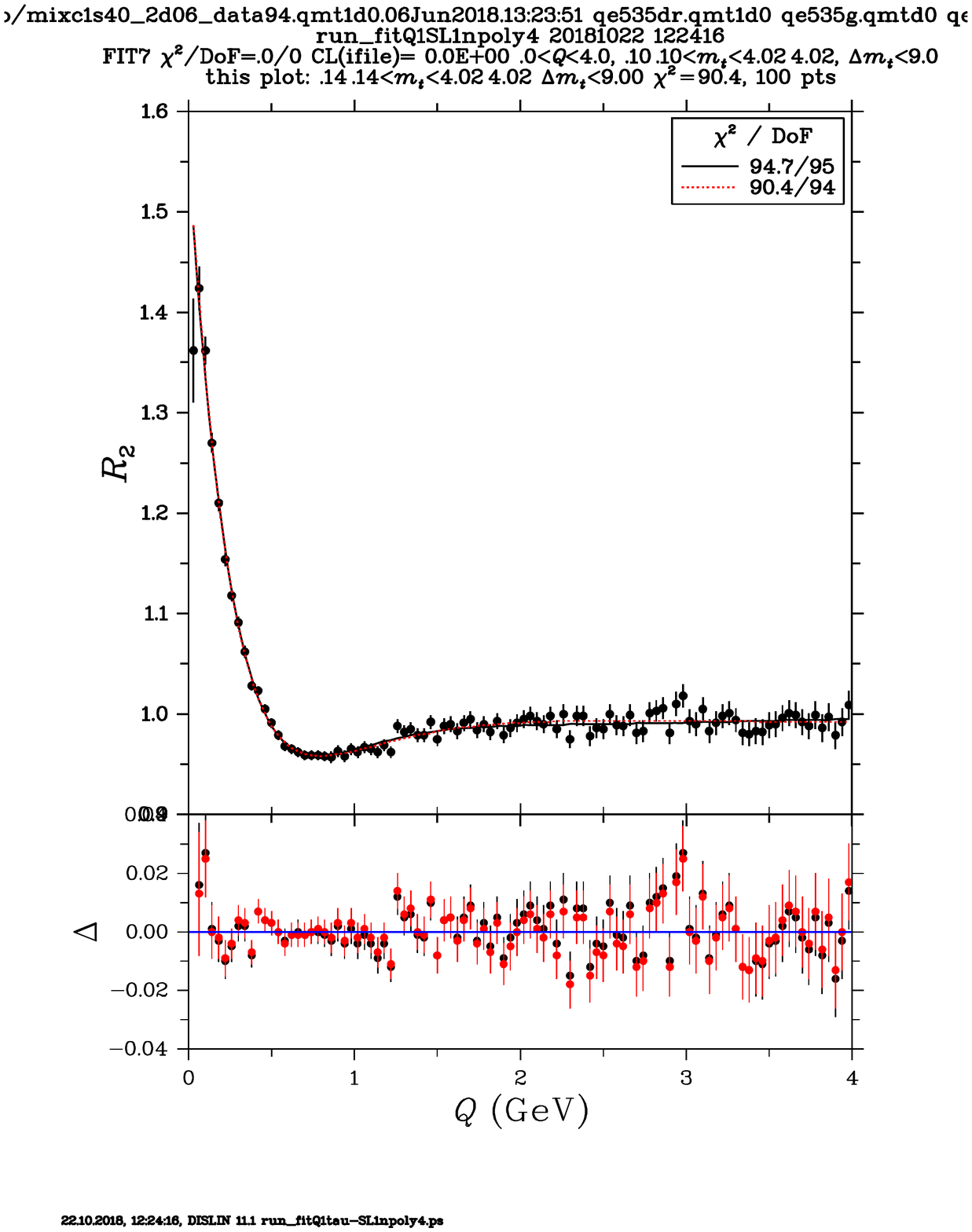}}
   \raisebox{75mm}{\hspace{-.8\textwidth}(a)\hspace{.43\textwidth}(b)}
        \caption{(a) Symmetric L\'evy fit for two-jet events up to orders 0 through 3 L\'evy polynomials;
                 (b) First-order L\'evy polynomial fit (black) for two-jet events compared to the
                 \taumodel\ fit (red).
        }
   \label{fig:R2expan}
\end{figure}

The symmetric L\'evy distribution can be expanded in terms of L\'evy polynomials~\cite{DeKock:WPCF2011,Csorgo:Odderon}, $l_i$,
which are orthonormal.
The resulting expression for $R_2$ is
\begin{equation} \label{eq:R2levypol}
         R_2(Q) = \gamma
                   \left[ 1+\lambda \exp\left(-\abs{rQ}^\alpha\right)
                             (1 + \sum c_i l_i )
                   \right]
                   \cdot\left(1+\epsilon Q\right) \;.
\end{equation}
Fits to the two-jet data are shown in \Fig{fig:R2expan}a for orders 0 through 3 of the L\'evy polynomials.
The order-0 fit (also shown in \Fig{fig:symlevy2jet}a) has a very poor \chisq, but the order-1 fit has a good \chisq.
Higher orders show only marginal further improvement.
 
The first-order symmetric L\'evy polynomial fit is shown together with the simplified \taumodel\ fit in
\Fig{fig:R2expan}b.  The \chisq\ of the L\'evy polynomial fit is slightly better than the \taumodel\ fit,
but the fit curves are nearly identical, what difference there is being mainly for $Q>1.5\,\GeV$.
 
\begin{figure}[h]
   \mbox{\includegraphics[width=.99\textwidth]{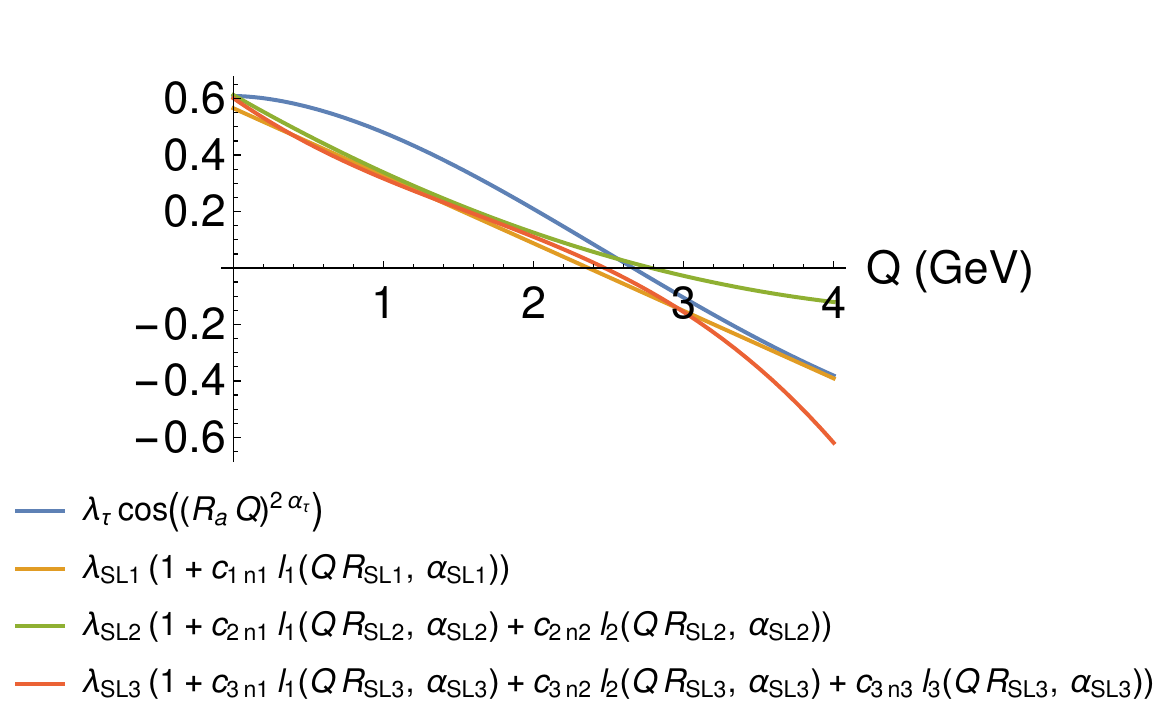}}
   \caption{Comparison of the cosine term of the \taumodel\ parametrization with the L\'evy polynomial term of the
            symmetric L\'evy parametrization.}
   \label{fig:cospolycompare}
\end{figure}

Comparing \Eqs{eq:asymlevR2} and (\ref{eq:R2levypol}), we see that in the symmetric L\'evy parametrization
the cosine of the \taumodel\ parametrization is replaced by the L\'evy polynomial expansion.
Also, in the exponential $2\alpha$ becomes simply $\alpha$.
\Fig{fig:cospolycompare} compares the cosine and the L\'evy polynomials.  We see a rather similar behaviour:
Both decrease more or less linearly with $Q$, which explains why both fit the data approximately equally well.
 
\subsection{\taumodel\ (asymmetric L\'evy)}
\begin{figure}[h]
   \mbox{\includegraphics[width=.45\textwidth,clip,viewport=56 87  518 680]{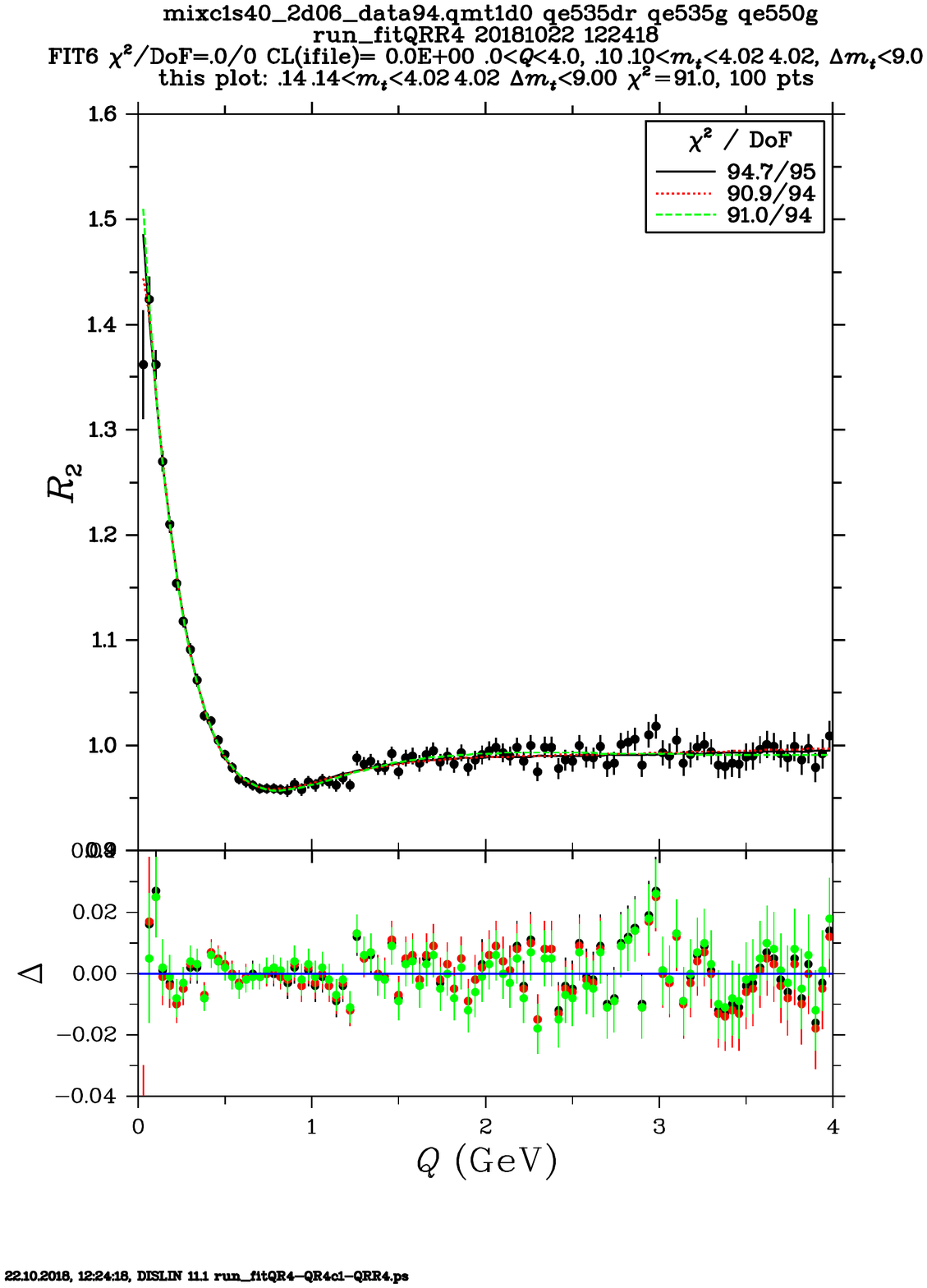}}
   \mbox{\includegraphics[width=.45\textwidth,clip,viewport=56 87  518 680]{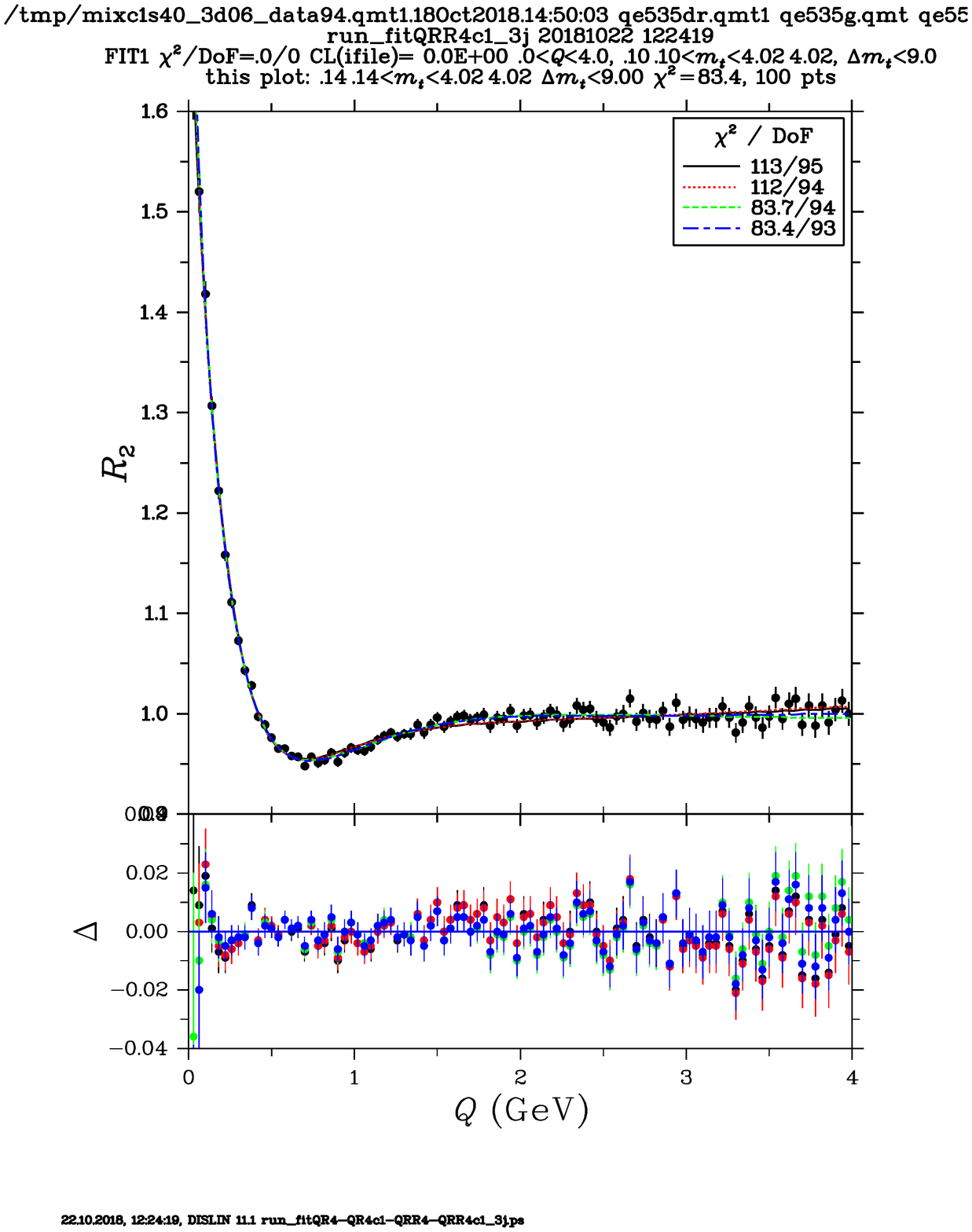}}
   \raisebox{75mm}{\hspace{-.8\textwidth}(a)\hspace{.43\textwidth}(b)}
        \caption{(a) Fits of the \taumodel\ (black), the order-1 expansion of the \taumodel\ (red), and the \taumodel\ with \Ra\
                     free (green), for two-jet events;
                 (b) Fits of the \taumodel\ (black), the order-1 expansion of the \taumodel\ (red), the \taumodel\ with \Ra\ free
                     (green), and the order-1 expansion of the \taumodel\ with \Ra\ free (blue), for three-jet events.
        }
   \label{fig:R2tauexpan}
\end{figure}

Lacking an orthogonal polynomial expansion for the asymmetric L\'evy distribution $H(\tau)$ of the \taumodel,
we use, motivated by the results of Ref.~\citen{DeKock:WPCF2011}, a derivative expansion:
\begin{eqnarray}
     R_2(Q) &=& {\gamma}
        \left[ 1+ \lambda \left\{
        \cos \left(\left(\Ra Q\right)^{2\alpha}\right)
             \exp\left(-\left(RQ\right)^{2\alpha}\right)
          \right. \right. \qquad\qquad\qquad  \nonumber \\
           & &  \left. \left.
             + \sum   c_n \frac{{\textrm d}^n\phantom{Q}}{{\textrm d}Q^n}
        \cos \left(\left(\Ra Q\right)^{2\alpha}\right)
             \exp\left(-\left(RQ\right)^{2\alpha}\right)
             \right\}
             \right]
             \cdot\left(1+ \epsilon Q\right)       \;.
\end{eqnarray}
We also consider letting \Ra\ be a free parameter rather than as defined in \Eq{eq:asymlevRaR}.
The results of these fits are shown in \Tab{tab:fits} and \Fig{fig:R2tauexpan}a.
The \chisq\ of the order-1 fit is, of course, smaller than that of the order-0 fit, as is the \Ra-free fit,
and the confidence levels are somewhat better.
To test the significance of the improvement in \chisq, we make use of the \chisq-difference.
For the order-0 and order-1 fits this is $\chisq_\text{diff}=94.7-90.9=3.8$, and the difference in the number of degrees of freedom
is 1.  The confidence level for a \chisq\ of 3.8 with 1 degree of freedom is 5.1\%.
A small value of this confidence level, say less than 5\%, would be grounds for rejecting the order-0 parametrization. (This
corresponds to the 95\% commonly used in making decisions.)
Thus we conclude that for two-jet events the order-0 fit is adequate.
 
\begin{table}[b]
\centering
\caption{Fit results of \taumodel\ parametrizations for two-jet events.}
\label{tab:fits}
   $ \begin{array}{lccc}
      \hline
              &\text{order 0}&\text{order 1}& \text{order 0, \Ra\ free} \\ 
      \hline
       \alpha & 0.44\pm0.01  & 0.43\pm0.01     & 0.41\pm0.02      \\ 
 R\ \text{(fm)}& 0.78\pm0.04  & 0.84\pm0.05     & 0.79\pm0.04     \\ 
 \Ra\ \text{(fm)}& -            & -             & 0.69\pm0.04     \\ 
       \lambda& 0.61\pm0.03  & 0.67\pm0.05     & 0.63\pm0.03      \\ 
       \gamma & 0.979\pm0.002& 0.979\pm0.002   & 0.988\pm0.005    \\ 
       \epsilon&0.005\pm0.001& 0.005\pm0.001   & 0.001\pm0.002    \\ 
        c_1   & -            & 0.0008\pm0.0005 & -                \\ 
      \hline
       \chisq/{\text{DoF}}\hspace{-2mm} &94.7/95 &90.9/94 &91.0/94\\ 
       \text{CL}& 49\%  & 57\% & 57\%  \\ 
     \hline
    \end{array} $
\end{table}
 
The order-0, \Ra-free fit also provides an adequate description, having a \chisq\ nearly identical to that of the order-1,
\Ra-constrained fit.
Note that the physical parameters ($\alpha$, $R$, $\lambda$) differ at most by about 1 standard deviation in going from
order-0 to order-1 or to \Ra\ free.  Thus conclusions based on these values, such as the reconstructions of the space-time picture
in Ref.~\citen{L3_levy:2011}, remain valid.

\section{Conclusions for two-jet events} \label{sect:2jetConclusions}
We have used expansions about the hypothesized form to test whether it provides an adequate description
of the data or is only a (poor) approximation.  In the latter case the shape of the expansion terms provide an indication of how
to modify the original parametrization.  For the symmetric L\'evy parametrization this showed that an approximately linearly
decreasing function of $Q$ is necessary, which in fact is what is provided by the \taumodel.
 
An expansion in the case of the \taumodel\ is found not to be significant, \ie, the one-sided L\'evy distribution of the \taumodel\
is adequate.

\section{Three-jet events} \label{sect:3jet}
For two-jet events hadronization occurs basically in 1+1 dimensions, which lead to the dependence of $R_2$ on
$\tau$, the longitudinal proper time and \mt, the transverse mass.
For three-jet events, the \Pq\Paq\Pg\ system no longer forms a linear system (in the overall centre of mass), but a planar one.
There is no event axis by which the transverse mass and longitudinal proper time are defined.
Therefore we might expect the \taumodel, as formulated for a two-jet system, to work less well.
 
The results of fits of the \taumodel\ and its first-order expansion, without and with \Ra\ a free parameter, are shown in
\Fig{fig:R2tauexpan}b and \Tab{tab:fits3j}.  Applying the \chisq-difference test to the order-0 and order-1 fits yields a
confidence level of 37\% for the case that \Ra\ is constrained and 58\% when \Ra\ is free.
 
However, the \chisq-difference between the order-0 \Ra\ constrained and free cases yields a confidence level of $6\cdot10^{-8}$.
Thus regarding \Ra\ as a free parameter does give significant improvement.
 
But it must be pointed out that the parameters \Ra\ and
$\tau_0$ are expected to be highly correlated. While for two-jet events it was found in Ref.~\citen{L3_levy:2011} that $\tau_0$ is
consistent with zero, such studies have not yet been performed for three-jet events.
Further investigation is ongoing.
Also, note that the value of $\alpha$ is significantly less for the fits with \Ra\ free than for those with \Ra\ constrained.
This was not the case for two-jet events.  This too requires additional investigation.

\begin{table}[t]
\centering
\caption{Fit results of \taumodel\ parametrizations for three-jet events.}
\label{tab:fits3j}
   $ \begin{array}{lcccc}
      \hline
              &\text{order 0}&\text{order 1}& \text{order 0, \Ra\ free} & \text{order 1, \Ra\ free} \\
      \hline
       \alpha & 0.42\pm0.01  & 0.42\pm0.01     & 0.35\pm0.01       & 0.35\pm0.01 \\
 R\ \text{(fm)}& 0.98\pm0.04  & 0.96\pm0.05     & 1.06\pm0.05      & 1.00\pm0.11 \\
 \Ra\ \text{(fm)}& -            & -             & 0.87\pm0.04      & 0.96\pm0.30 \\
       \lambda& 0.84\pm0.04  & 0.81\pm0.05     & 0.92\pm0.05       & 0.65\pm0.49 \\
       \gamma & 0.977\pm0.001& 0.977\pm0.001   & 0.997\pm0.005     & 0.994\pm0.007 \\
       \epsilon&0.008\pm0.001& 0.008\pm0.001   & 0.0003\pm0.0017    & 0.001\pm0.002 \\
        c_1   & -            &-0.0005\pm0.0005 & -                 &-0.04 \pm0.08  \\
      \hline
       \chisq/{\text{DoF}}\hspace{-2mm} &113.2/95 &112.4/94 &83.7/94  & 83.4/93 \\
       \text{CL}& 10\%  &  9\% & 77\%    & 75\% \\
     \hline
    \end{array} $
\end{table}

\section{Conclusions for three-jet events} \label{sect:3jetConclusions}
As for the two-jet case, expansion of the \taumodel\ expression does not lead to significant improvement in the fits.
This validates the use of an asymmetric L\'evy distribution for the longitudinal proper time.
 
However, significant improvement of the fit is obtained by letting \Ra\ be a free parameter. \ie,
by lessening the connection of the simplified \taumodel\ between the BEC peak and the antisymmetric dip.
Whether letting $\tau_0$ also be a free parameter would also give significant improvement is the subject of ongoing investigation,
as is the question whether $\alpha$ decreases as the number of jets increases.
 
\begin{acknowledgement}
S.~L\"ok\"os greatfully acknowledges the support from the ERASMUS project of the EU and thanks W. Metzger for
his kind hospitality at the Radboud University Nijmegen, The Netherlands.
This research was partially supported by the  NKTIH FK 123842 and FK 123 959 grants (Hungary),
by the Circles of Knowledge Club (Hungary) and by  the  EFOP 3.6.1-16-2016-00001 project (Hungary).
\end{acknowledgement}

 
\newpage
\hyphenation{Post-Script Sprin-ger}

\end{document}